\documentclass[prl,showpacs,onecolumn]{revtex4}

\usepackage{epsfig}
\usepackage{color}

\usepackage{graphicx}

\usepackage{amssymb,amsfonts,amsmath}

\begin{document}

\title{Population genetics in compressible flows}

\author{Simone Pigolotti$^{1,2}$, Roberto Benzi$^3$, Mogens
  H. Jensen$^1$ and David R. Nelson$^4$} \affiliation{$^1$The Niels
  Bohr Institut, Blegdamsvej 17, DK-2100 Copenhagen, Denmark. $^2$
  Dept. de Fisica i Eng. Nuclear, Universitat Politecnica de Catalunya
  Edif. GAIA, Rambla Sant Nebridi s/n, 08222 Terrassa, Barcelona,
  Spain. $^3$ Dipartimento di Fisica, Universita' di Roma ``Tor
  Vergata'' and INFN, via della Ricerca Scientifica 1, 00133 Roma,
  Italy. $^4$ Lyman Laboratory of Physics, Harvard University,
  Cambridge, MA 02138, USA}

\date{today}

\begin{abstract} 

  We study competition between two biological species advected by a
  compressible velocity field.  Individuals are treated as discrete
  Lagrangian particles that reproduce or die in a density-dependent
  fashion.  In the absence of a velocity field and fitness advantage,
  number fluctuations lead to a coarsening dynamics typical of the
  stochastic Fisher equation.  We then study three examples of
  compressible advecting fields: a shell model of turbulence, a
  sinusoidal velocity field and a linear velocity sink. In all cases,
  advection leads to a striking drop in the fixation time, as well as
  a large reduction in the global carrying capacity.  Despite
  localization on convergence zones, one species goes extinct much
  more rapidly than in well-mixed populations.  For a weak harmonic
  potential, one finds a bimodal distribution of fixation times. The
  long-lived states in this case are demixed configurations with a
  single boundary, whose location depends on the fitness advantage.

\end{abstract}


\maketitle

Challenging problems arise when spatial migrations of species are
combined with population genetics.  The population dynamics of a
single species expanding into new territory was first studied in the
pioneering works of Fisher, Kolmogorov, Petrovsky and Piscounov (FKPP)
\cite{1,2,3}. Later, Kimura and Weiss studied individual-based
counterparts of the FKPP equation \cite{6}, revealing the important
role of number fluctuations. In particular, stochasticity is
inevitable at a frontier, where the population size is small and the
discrete nature of the individuals becomes essential.  Depending on
the parameter values, fluctuations can produce radical changes with
respect to the deterministic predictions \cite{3,4}.  If $f(x,t)$ is
the population fraction of, say, a mutant species and $1-f(x,t)$ that
of the wild type, the stochastic FKPP equation reads in one dimension
\cite{8}: 
\begin{equation}\label{fkpp}
\partial_t f(x,t) = D\partial^2_x f + s f(1-f) +\sqrt{D_g f(1-f)}\xi (x,t)
\end{equation}
where $D$ is the spatial diffusion constant, $D_g$ is the genetic
diffusion constant (inversely proportional to the local population
size $N_l$), $s$ is the genetic advantage of the mutant and
$\xi=\xi(x,t)$ is a Gaussian noise, delta-correlated in time and space
that must be interpreted using Ito calculus \cite{8,9}.

However, many species, from the distant past \cite{11} up to the
present, have competed in liquid environments, such as lakes, rivers
and oceans. Interesting new phenomena arise when population dynamics
is coupled to hydrodynamic flows \cite{5}.  For example, satellite
observations of chlorophyll concentrations have identified long lived,
segregated patches of marine microorganisms off the eastern coast of
the southern tip of South America \cite{12}, where species domains are
largely determined by the tangential velocity field obtained from
satellite altimetry.  In cases such as these, the dynamics takes place
in the presence of advecting flows, some of them at high Reynolds
numbers \cite{13}.  It is often appropriate to consider a {\em
  compressible} velocity field, both because of inertial effects
\cite{14} associated with fairly large microorganisms (diameter
5-500$\mu$m), and because photosynthetic bacteria and plankton often
control their buoyancy to stay close to the ocean surface \cite{15}.
In the latter case, the coarse-grained velocity field advecting the
microorganisms will contain a compressible component to account for
downwellings \cite{16}.  Recent works studied the dynamics of a
single population in the presence of a compressible turbulent velocity
field in one and two dimensions \cite{17,18}.  Here, the interplay
between turbulent dynamics and population growth leads to
“quasilocation” on convergence zones and a remarkable reduction in the
carrying capacity.

The study of {\em competition} in a hydrodynamics context, where both
a compressible velocity field and stochasticity due to finite
population sizes are present, calls for a nontrivial generalization of
Eq. (\ref{fkpp}). One complication is that, because of compressibility,
the sum of the concentrations of the two species is no longer
invariant during the dynamics. Thus, we must clarify the definition of
$f(x,t)$, the fraction of one particular species. A biologically
important issue arises from overshooting: the density $f(x,t)$ can
exceed unity near velocity sinks, resulting in an unphysical imaginary
noise amplitude in Eq. (\ref{fkpp}).

In this Letter, we overcome these problems by introducing a new model,
designed to explore, for the first time, how compressible velocity
flows affect the competition between two different species which can
die and reproduce via cell division.  In absence of an advecting
field, the model reproduces the coarsening dynamics of spatial
population genetics, as shown for one spatial dimension in Fig. 1a.
Of particular importance in population genetics is the {\em fixation
  time} $\tau_f$, defined as the average time needed for one of the
two species to reach extinction. The value of $\tau_f$ is proportional
to the total population size in mean field approximation
(i.e. assuming uniform spatial mixing) and grows to $L^2/D$ for a one
dimensional system \cite{8}. Conversely, Fig. 1b shows a typical
simulation with our model using a high Reynold number synthetic
velocity field extracted from a shell model of turbulence \cite{17a}.
In striking contrast to the result without advecting flows (Fig. 1a
and Ref. \cite{8}), one now finds competitions carried out on
transient but persistent sinks in the velocity field. We find that
turbulence not only compresses and reduces the overall population
density, it also greatly reduces the fixation time $\tau_f$. Despite
quasi-localization onto velocity sinks, the typical fixation time is
much shorter than if particles were well mixed.  We have also
considered a compressible sinusoidal velocity field and a confining
sink arising from a linear velocity profile.  Remarkably, we find that
these low Reynolds number compressible flows can {\it also}
dramatically reduce fixation times. In the following, we investigate
this phenomenon and propose a phenomenological explanation.

Eq. (\ref{fkpp}) is derived by assuming a fixed total concentration of
individuals, so that if $f(x,t)$ is the concentration of one species,
the other will have a concentration exactly equal to $1-f(x,t)$. To
describe cases in which the total concentration can change due to an
advecting flow, we introduce an off-lattice model in which two
different organisms, $A$ and $B$, advect and diffuse in space, while
undergoing duplication (i.e. cell division) and density-dependent
annihilation (death). Specifically, we implement the following
reactions: each particle of species $i=A,B$ duplicates with rate
$\mu_i$ and annihilates with a rate $\bar{\mu}_i \widehat{n}_i$, where
$\widehat{n}_i$ is the number of neighboring particles (of both types)
in an interaction range $\delta\propto1/N$, and $N$ is the total number of
organisms that can be accomodated in the unit interval with total
density $c_A+c_B=1$.  To reduce the number of parameters, we set
$\bar{\mu}_A=\bar{\mu}_b=\mu_B=\mu$, but take $\mu_A=\mu(1+s)$ to allow
the possibility of a selective advantage (faster reproduction rate) of
species $A$. We will start by analyzing in depth the neutral case
$s=0$ and consider the effect of $s>0$ in the end of the Letter.
In one dimension, our macroscopic coupled equations for the densities
$c_A(x,t)$ and $c_B(x,t)$ of individuals of type $A$ and $B$ in an
advecting field $v(x,t)$ read

\begin{eqnarray}\label{model}
\partial_t c_A\!\!&=\!\!&\!
-\partial_x(vc_A)\!+\!D\partial^2_xc_A\!+\!\mu c_A(1\!+\!s\!-\!c_A\!-\!c_B)\!+\!
\sigma_A\xi
\nonumber\\
\partial_t c_B\!\!&=\!\!&\!
-\partial_x(vc_B)\!+\!D\partial^2_xc_B\!+\!\mu c_B(1\!-\!c_A\!-\!c_B)\!+\!
\sigma_B\xi'\ 
\end{eqnarray}
with $\sigma_A=\sqrt{\mu c_A(1+s+c_A+c_B)/N}$ and $\sigma_B=\sqrt{\mu
  c_B(1+c_A+c_B)/N}$. As discussed in \cite{suppl}, for an interval of
size $L$, let $N^{-1}\rightarrow L/N=\delta/N_l$, where $N_l$ is the
local carrying capacity of a region of size $\delta$.  $\xi(x,t)$ and
$\xi'(x,t)$ are independent delta-correlated noise sources with an
Ito-calculus interpretation as in Eq. (\ref{fkpp}), and we have
rescaled the particle densities so that the local carrying capacity is
$1$ when $v=s=0$. Although $s=0$ may seem
nongeneric in the context of dynamical systems theory, this is a case
of particular interest in population genetics, where it corresponds to
the neutral theory of Kimura \cite{6}.

\begin{figure}[htb]
\begin{center}
\includegraphics[width=8.4cm]{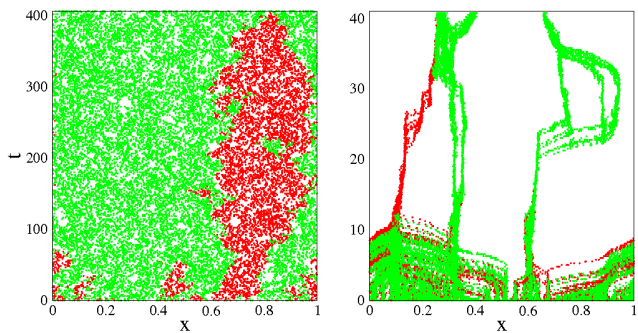}
\caption{Space-time plot of the off-lattice particle model, (a) no
  advecting velocity field and (b) with a compressible turbulent
  velocity field. Simulations are run until fixation (disappearance of
  one of the two species); note the reduced carrying capacity and the
  much faster fixation time in (b). Parameters values: $N=10^3$, $D=
  10^{-4}$, $\mu=1$. Parameters of the shell model are as in
  \cite{17}. \label{fig1} }
\end{center}
\end{figure}

The above equations follow from a microscopic master equation via the
Kramers-Moyal method \cite{suppl}, at this order equivalent to Van
Kampen inverse system size expansion \cite{20}.  Multi-species
versions of the FKPP equations have been already considered in Refs
\cite{21}, but not in the presence of an advecting field or number
fluctuations. 

We first simulated the particle model directly for $v=s=0$, finding a
dynamics similar to that of the stochastic FKPP equation,
Eq. (\ref{fkpp}), for $s=0$, see Fig.(\ref{fig1}a). In this simple
limit, our model can be considered as a grandcanonical generalization
of Eq (\ref{fkpp}), where the total density of individuals $c_A + c_B$
is now allowed to fluctuate around an average value $1$. Details of
the numerical implementation and the role of density fluctuations are
presented in \cite{suppl}.  Hereafter, we fix the following
parameters: $N=10^3$, $D= 10^{-4}$, $\mu=1$ and $L=1$ where $L$ is the
one dimensional domain endowed with periodic boundary conditions. With
these parameters, $\tau_f$ would be $\sim 10^4$ for the one
dimensional FKKP equation, and $\sim 10^3$ for the well-mixed case.

Introducing a compressible velocity field $v(x,t)$, as shown in
Fig.(\ref{fig1}b), leads to radically different dynamics.  Individuals
tend to concentrate at sinks in the velocity field. Further,
competition is enhanced and the total number of individuals $n(t)$
present at time $t$ is on average significantly smaller than the total
carrying capacity $N$.  We implemented three different velocity
fields: 1) a velocity field $v(x,t)$ generated by a shell model of
compressible turbulence \cite{17a,17}, reproducing the power spectrum
of a high Reynolds number turbulent velocity field of forcing
intensity $F$ \cite{suppl}, 2) a static sine wave, $v(x)=F \sin(2\pi
x)$, and 3) a linear velocity profile, $v(x)=-k x$, confining
particles near the origin. Periodic boundary conditions on the unit
interval are used in cases (1) and (2), while in case (3) the system
size is so large that particles never reach the boundaries.

\begin{figure}[htb]
\begin{center}
\includegraphics[width=8.4cm]{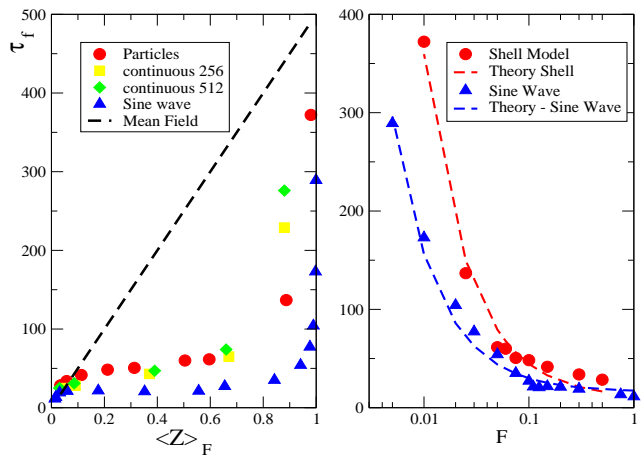}
\caption{Average fixation time $\tau_f$ for neutral competition in
  compressible turbulence and sine wave advection, as a function of
  (left) the reduced carrying capacity $\langle Z\rangle_F$ and
  (right) forcing intensity $F$ (smaller $\langle Z\rangle_F$ in the
  left panel corresponds to larger forcing in the right panel). In (a)
  red circles and blue triangles are particle simulations. Other
  symbols denote simulations of the continuum equations with different
  resolutions on the unit interval. The black dashed line is the mean
  field prediction, $\tau_f=N\langle Z\rangle_F/2$. In (b), only
  particle simulations are shown and dashed lines are the theoretical
  prediction $\tau_f=\tau_0+c/F$ based on boundary domains, with
  fitted parameters $\tau_0=9.5$, $c=3.5$ in the case of the shell
  model and $\tau_0=16$, $c=1.4$ in the case of the sine
  wave.\label{fig2}}
\end{center}
\end{figure}

Fig.(\ref{fig2}) shows the average fixation time $\tau_f$ for $s=0$ in
the first two cases, at varying the intensity $F$ of advection.  In
the left panel, we plot the fixation times as a function of the time
averaged reduced carrying capacity $\langle Z \rangle$, where
$Z(t)=n(t)/N$ is the carrying capacity reduction, i.e. the ratio
between the actual number of particles and the average number of
particles $N$ observed in absence of the velocity field.  Plotting
vs. $\langle Z \rangle_F$ allows comparisons with the mean field
prediction, $\tau_f=2N \langle Z\rangle_F/\mu$, valid for well mixed
systems (black dashed line) \cite{8}.  For the shell model, we include
simulations of the macroscopic equations (\ref{model}) with different
resolutions ($256$ and $512$ lattice sites on the unit interval),
obtaining always similar results for $\tau_f$
vs. $\langle Z \rangle_F$.

In all cases, the presence of a spatially varying velocity field leads
to a dramatic reduction of $\tau_f$, compared to the mean field
theory. The fixation time drops abruptly as soon as $\langle Z \rangle
<1$, even for very small $F$. Simulations suggest a singular limit of
zero intensity ($F \sim 0$) of the velocity field (and consequently
$\langle Z\rangle \rightarrow 1$ in Fig. 2), as we discuss later.

These observations can be understood by assuming that global fixation
is determined by coalescence of allele boundaries like those shown in
Fig. (\ref{fig1}a). Although in the stochastic FKPP with $s=0$ the
boundaries perform a random walk \cite{8}, here they are also advected
by the velocity field.  In particular, the average position of one
boundary $x$, neglecting diffusion and number fluctuations, will
simply evolve according $\dot{x}=v(x,t)$. The fixation time is then
determined by the time needed for the boundaries to reach the center
of a sink and annihilate. Because periodic boundary conditions are
imposed, the number of interfaces is necessarily even, so all of them
annihilate pairwise. Upon expanding the velocity field around a sink
position $x_0$, $v(x)=k(x_0-x)$, we find a fixation time $\sim
k^{-1}$. Hence, we model the average time to fixation $\tau_f$ vs. the
velocity field intensity as $\tau_f$=$\tau_0+ck^{-1}$, where $\tau_0$
represents the typical time to reach fixation at large strain rate $k$
and $c$ is an dimensionless constant of order unity. This
phenomenological theory describes well the data of Fig. \ref{fig2},
right panel (dot-dashed lines), where in the turbulent case we take
the forcing $F$ as a proxy for $k$ ($k=2\pi F$ for the sine wave and
$k=\sqrt{\langle(\nabla v)^2\rangle}\propto F$ for turbulence).

\begin{figure}[htb]
\begin{center}
\includegraphics[width=8.4cm]{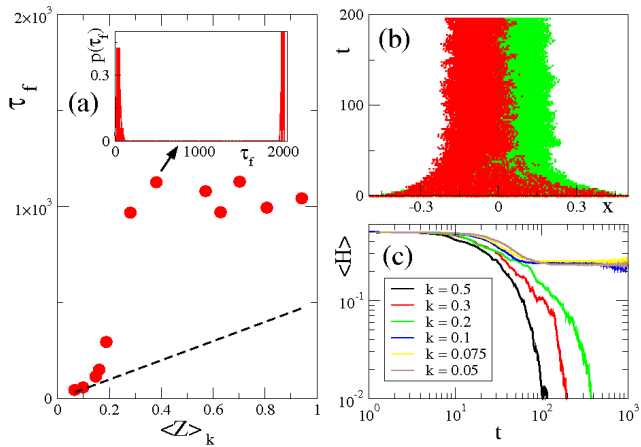}
\caption{(a) Average fixation time $\tau_f$ in the presence of
  a linear converging flow $v=-k x$, as a function of the reduced carrying
  capacity $\langle Z \rangle_k$.  Inset on left shows the
  distribution of fixation times in the case $k=0.075$ and $\langle Z
  \rangle_k\approx 0.38$, the right peak representing all realizations
  with fixation times $t>2000$. (b) Space-time plot of a realization
  with a very long fixation time. (c) Average heterozigosity as a
  function of time for different values of $k$.
  \label{fig3}}
\end{center}
\end{figure}

We now study particles subject to a converging linear velocity,
$v(x)=-k x$. When $k$ is large (and $\langle Z\rangle_k$ is small),
the fixation time is comparable to the mean field prediction. However,
as the forcing decreases, it becomes much {\it larger} than the mean
field prediction (Fig. \ref{fig3}a, dashed line). In this regime, the
fixation time probability distribution is bimodal (Fig. \ref{fig3}a,
inset): roughly half of the realizations have a short fixation time
(faster than mean field), while the other half maintain coexistence
for an extended period. The space-time evolution of these
configurations in Fig. \ref{fig3}b reveals that the two mutant species
are {\it demixed}, one on the left and one on the right of the
sink. Correspondingly, the average total heterozigosity $\langle H(t)
\rangle$, defined as the probability of two random individuals to be
of different type \cite{8}, decays exponentially for very large $k$
(consistent with mean field theory \cite{8}), but tends to a constant,
non-zero value when smaller values of $k$ are chosen (Fig
\ref{fig3}c).

This remarkable behavior arises because of the different boundary
conditions for the linear sink, such that the number of domain
boundaries is no longer always even. Hence, the initial
number of boundaries will be odd approximately half the time for
random initial conditions, leading to the single surviving boundary as
shown on the right of Fig. \ref{fig3}a. This configuration corresponds
to a stable stationary solution of the deterministic version of Eqs
\ref{model} with $s=0$. We expect that this demixed solution, with
$c_A(x)$ and $c_B(x)$ nonzero on opposite sides of the sink, should
have a lifetime (inaccessible in numerical simulations) which
grows exponentially with $N$ \cite{9,20}. 

\begin{figure}[htb]
\begin{center}
\includegraphics[width=8.4cm]{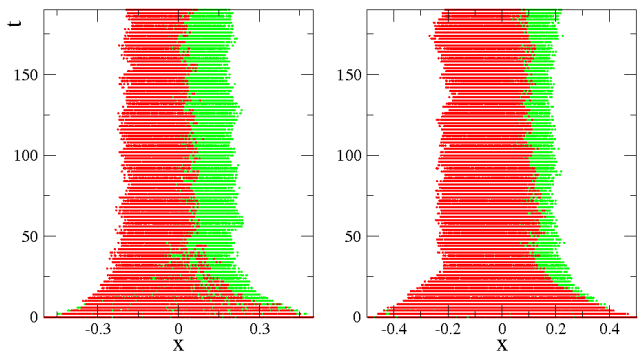}
\caption{ Coexistence of two species advected by a linear sink
  $k=0.075$, (left) species are neutral as in Fig. 3, (right) Red
  reproduces 30\% faster. Notice the shift in the position of the
  boundary.\label{fig4}}
\end{center}
\end{figure}

Finally, what happens for a selective advantage $s>0$?
Fig. \ref{fig4} shows a comparison between the neutral case of
Fig. \ref{fig3} and a simulation in which red particles reproduce
$30\%$ faster ($s=0.3$). Even with such a large selective advantage,
genetic interfaces are still present. In this case, we estimate the
shift in the interface as $\delta x=k^{-1}2\sqrt{Ds}$, by equating the
outward Fisher genetic wave velocity $v_g=2\sqrt{Ds}$ \cite{1} with
the inward advection velocity at distance $\delta x$ from the origin.

To conclude, we have introduced a new model which allows studies of
how compressible advection affects Darwinian competition.  A
compressible flow leads to a radically new scenario, with the fixation
time being largely determined by how species boundaries collapse into
the sinks of the velocity field, rather than by diffusive
annihilation.  Our phenomenological theory suggests a significative
effect even for very weak compressible fields. Compressible advection
becomes irrelevant for the fixation time only when the time to drift
into the sinks (of order $t\sim k^{-1}$) is much larger than the
diffusion time (of order $L^2D^{-1}$, where $L$ is a typical linear
size of the population). An obvious application of the model is the
study of compressible flows in higher space dimensions, including
flows relevant to biological oceanography of photosynthetic organisms
near the water surface. Recent advances in experimental techniques
could also lead to tests of the one dimensional results.  Motility of
bacteria has been studied in microstructures of diameter comparable or
even smaller to that of the organisms \cite{mannik}. In this setting,
a compressible flow could be created by pumping liquid nutrient from
the two ends of the tube and extracting it from the center via a
semipermeable membrane. Another possibility could be to study floating
microorganisms \cite{15,rainey} and make use of the compressible
effects caused by the vertical component of a convecting velocity
field \cite{16}.

\begin{acknowledgments}
  We are grateful to F. Toschi and P. Perlekar for interesting
  discussions. Work by SP and MHJ was supported by the Danish National
  Research Foundation through the Center for Models of Life. Support
  for DRN was provided by the National Science Foundation in part
  through Grant No. DMR-1005289 and by the Harvard Materials Research
  Science and Engineering Center through NSF Grant No. DMR-0820484.
\end{acknowledgments}

\section{SUPPLEMENTARY MATERIALS}

In this note we present the technical derivations of the results in
the Letter ``Population Genetics in compressible flows'' (``main
text''). First, we describes the derivation of the
macroscopic equations for the particle model introduced in the main
text. Then, we describe details of the continuum
simulations and of the shell model of turbulence. Finally, we
discuss the relation between the model introduced in the main text
and the simpler stochastic Fisher-Kolmogorov-Petrovsky-Piskounov
(FKPP) equation (Eq. (1) of the main text).

\section{Macroscopic equations for the two species model}

To derive the mean field version of the model discussed in the Letter.
we consider first position-independent quantites $n_A$ and $n_B$ which
represent the number of particles of species $A$ and $B$
respectively. This ``zero-dimensional'' limit corresponds to a single
well mixed ``deme'' or island in a stepping stone model. The birth and
death processes  are characterized by the following
transition rates:

\begin{eqnarray}\label{mrates}
  W_A(+1,n_A,n_B)&=&\mu n_A\nonumber\\
  W_A(-1,n_A,n_B)&=&\tilde{\mu}n_A(n_A+n_B)\nonumber\\
  W_B(+1,n_A,n_B)&=&\mu n_B\nonumber\\
  W_B(-1,n_A,n_B)&=&\tilde{\mu}n_B(n_A+n_B)
\end{eqnarray}
where $W_A(+1,n_A,n_B)$ is the probability per unit time of population
$A$ to increase its size by one unit when the total populations
are $n_A$ and $n_B$, and similarly for $W_A(-1,n_A,n_B)$ and $W_B(\pm
1,n_A,n_B)$.

There exist several methods to derive macroscopic equations for
particle models \cite{9,risken,20}, most popular
being the Van Kampen system size expansion and the Kramers-Moyal
expansion. While the former is more rigorous when studying problems
beyond the linear noise approximation, the latter is normally used
because it leads to more transparent expressions. Moreover, it can be
shown (see e.g. \cite{9}, page 251) that at the order of the
Fokker Planck equation the two methods are completely equivalent.  The
Kramers-Moyal method works as follows: consider a master equation, and
call $W(\Delta n, n)$ the transition rate from $n$ particles to
$n+\Delta n$ particles. When $n$ is typically large and $\Delta n$
relatively small ($\pm 1$ in the case of birth-death process
considered here) a Taylor series expansion of the master equation in
$\Delta n$ for the probability $P(n,t)$ of having $n$ particles at time
$t$ is appropriate:
\begin{equation}\label{KM1}
  \partial_t P(n,t)=\sum\limits_{j=1}^{\infty} 
\frac{(-1)^j}{j!}\partial^j_n[\alpha_j(n)P(n,t)]
\end{equation}
with
\begin{equation}\label{KM2}
\alpha_j(n)=\int d\Delta n \ (\Delta n)^j W(\Delta n,n).
\end{equation}
Truncating the Taylor expansion at the second order in $j$ leads to a
Fokker-Planck equation.

It is straightforward to apply the above procedure to two species $A$
and $B$ and obtain the zero-dimensional Fokker-Planck equation for the
joint probability distribution function $P(n_A,n_B,t)$. Substituting
the expression of the transition rates of Eq. (\ref{mrates}) into
(\ref{KM1}- \ref{KM2}) leads to:
\begin{eqnarray}\label{meanfieldFP}
\partial_t P(n_A,n_B,t)=-\partial_A[n_A(\mu-\tilde{\mu}n_A-\tilde{\mu}n_B)P]
-\partial_B[n_B(\mu-\tilde{\mu}n_A-\tilde{\mu}n_B)P]+\nonumber\\
+\frac{1}{2}\partial_A^2[n_A(\mu+\tilde{\mu}n_A+\tilde{\mu}n_B)P]
+\frac{1}{2}\partial_B^2[n_B(\mu+\tilde{\mu}n_A+\tilde{\mu}n_B)P],
\end{eqnarray}
corresponding to the stochastic differential equations:
\begin{eqnarray}\label{meanfieldL}
\frac{dn_A(t)}{dt}=n_A(\mu-\tilde{\mu}n_A-\tilde{\mu}n_B)
+\sqrt{n_A(\mu+\tilde{\mu}n_A+\tilde{\mu}n_B)}\xi(t)\nonumber\\
\frac{dn_B(t)}{dt}=n_B(\mu-\tilde{\mu}n_A-\tilde{\mu}n_B)
+\sqrt{n_B(\mu+\tilde{\mu}n_A+\tilde{\mu}n_B)}\xi'(t)
\end{eqnarray}
where $\xi(t)$ and $\xi'(t)$ are delta-correlated Gaussian processes
in time with unit amplitude,
e.g. $\langle\xi(t_1)\xi(t_2)\rangle=\delta(t_1-t_2)$ and are
uncorrelated with each other,
$\langle\xi(t_1)\xi'(t_2)\rangle=0$ . For consistency we
have to adopt the Ito prescription for interpreting the nonlinear noise term
\cite{9}.

From Eq. (\ref{meanfieldL}) we see that typical values of $n_A+n_B$
near the steady state are of order $N=\mu/\tilde{\mu}$. Upon defining
the rescaled particle densities
\begin{equation}
c_A=\frac{n_A}{N}=\frac{\tilde{\mu}}{\mu}n_A,
\quad c_B=\frac{n_B}{N}=\frac{\tilde{\mu}}{\mu}n_B ,
\end{equation}
we obtain
\begin{eqnarray}\label{zerodmodel}
\frac{dc_A(t)}{dt}=\mu c_A(1-c_A-c_B)+
\sqrt{\frac{\mu c_A(1+c_A+c_B)}{N}}\xi(t)\nonumber\\
\frac{dc_B(t)}{dt}=\mu c_B(1-c_A-c_B)+
\sqrt{\frac{\mu c_B(1+c_A+c_B)}{N}}\xi'(t).
\end{eqnarray}
These coupled Langevin equations approximate well the master equation
when $N=\mu/\tilde{\mu}$ is large. If $N$ is not large, the noise cannot be
approximated by a Gaussian noise.

The model is generalized to include advection and diffusion in space
as follows: Particles belonging to the two species 
advect and diffuse in space in a Lagrangian way, i.e. the position $x$
of a given particle evolves according to:
\begin{equation}
\frac{dx}{dt}=v(x)+\sqrt{2D}\zeta(t),
\end{equation}
where $\zeta(t)$ is a unit amplitude Brownian noise source that
implements diffusion with diffusion constant $D$.  At each time step
$dt$, a given particle duplicates with probability $\mu dt$ or
annihilates due to competition with rate
$\tilde{\mu}(\hat{n}_A+\hat{n}_B) dt$, where $\hat{n}_{A,B}$ are now
the number of particles of type $A,B$ in a region of size $\delta$
centered in the particle itself. 

The macroscopic equations in the spatial case can be derived using
arguments similar to the well mixed case. We assume that binary
reactions can occur only when both particles are within a length
$\delta$, and also take this as the size of our discretization volume.
Although the total particle number in this region fluctuates, $\delta$
plays a role similar to an island spacing in a stepping stone model.
We first write the discretized stochastic differential equation for
the $n_A^i$ and $n_B^i$ , the number of particles in the $i^{th}$ cell
of size $\delta$:
\begin{eqnarray}
  \frac{dn_A^i}{dt}=n_A^i(\mu-\tilde{\mu}n_A^i-\tilde{\mu}n_B^i)
+\mathrm{diffusion/advection}
  +\sqrt{n_A^i(\mu+\tilde{\mu}n_A^i+\tilde{\mu}n_B^i)}\xi^i_A
  \nonumber\\
  \frac{dn_B^i}{dt}=n_B(\mu-\tilde{\mu}n_A^i-\tilde{\mu}n_B^i)
+\mathrm{ diffusion/advection}
  +\sqrt{n_B^i(\mu+\tilde{\mu}n_A^i+\tilde{\mu}n_B^i)}\xi^i_B
\end{eqnarray}
where the noise terms now obey $\langle \xi^i_{k}(t)\xi^j_{k'}(t')\rangle
=\delta(t-t')\delta_{j,j'}\delta_{k,k'}$. 

Here, we assume the ration $N_l=\mu/\tilde{\mu}$ is adjusted to match
the overall carrying capacity of the $i^{th}$ cell, and that $\delta$ is big
enough to insure the local carrying capacity $N_l\gg 1$.  When $N_l$
is sufficiently large, the diffusion and advection terms enter only in
the deterministic part of our equations. 

There are two possible choices for passing to the concentration equations:
\begin{enumerate}
\item Change variables to the integrated dimensionless concentration
  inside each cell: $c_{A,B}^i=n^i_{A,B}/N_l$.
\item Change variables to a normalized real concentration (number of
  particles per unit volume): $c_{A,B}^i=n^i_{A.B}/(N\delta^d)$ in
  $d$ dimensions (we focus here on $d=1$).
\end{enumerate}
In the first case, one is left with a discretized set of well defined
Langevin equations, a natural generalization of the discrete stepping
stone model with dimensionless species concentrations. 
 The first procedure, after setting as before
$N_l\tilde{\mu}=\mu$, results in:
\begin{eqnarray}
  \frac{dc_A^i}{dt}&=&\mu c_A^i(1-c_A^i-c_B^i)
+\mathrm{diffusion/advection}
  +\sqrt{\frac{\mu c_A^i(1+c_A^i+c_B^i)}{N_l}}\xi^i_A
  \nonumber\\
  \frac{dc_B^i}{dt}&=&\mu c_B^i(1-c_A^i-c_B^i)
+\mathrm{ diffusion/advection}
  +\sqrt{\frac{\mu c_B^i(1+c_A^i+c_B^i)}{N_l}}\xi^i_B
\end{eqnarray}
which is a set of well defined stochastic differential equations,
valid for a particular lattice spacing $\delta$.

The second procedure, however, leads to a more convenient passage to
the continuum limit,

\begin{eqnarray}\label{spdiscr}
  \frac{dc_A^i}{dt}&=&c_A^i[\mu-\tilde{\mu}N\delta(c_A^i-c_B^i)]
+\mathrm{diffusion/advection}
  +\sqrt{\frac{c_A^i[\mu+\tilde{\mu}N\delta(c_A^i+c_B^i)]}{N\delta}}\xi^i_A
  \nonumber\\
  \frac{dc_B^i}{dt}&=&c_B^i[\mu-\tilde{\mu}N\delta(c_A^i-c_B^i)]
+\mathrm{ diffusion/advection}
  +\sqrt{\frac{c_B^i[\mu+\tilde{\mu}N\delta(c_A^i+c_B^i)]}{N\delta}}\xi^i_B
\end{eqnarray}

Starting with Eq.(\ref{spdiscr}), a formal way to achieve the
continuous limit is to set first $\tilde{\mu}N\delta=\mu$ by
rescaling the densities. This choice corresponds to an average number
of particles equal to $N_l$ in a cell in the absence of fluid flow
when the total density $c_A+c_B$ is constant and equal to $1$. Then,
we take the continuum limit at fixed $\mu$, leading to:
\begin{eqnarray}\label{model1}
\partial_t c_A=
-\partial_x(vc_A)+D\partial^2_xc_A+\mu c_A(1-c_A-c_B)+
\sigma_A\xi&
\\
\label{model2}
\partial_t c_B=
-\partial_x(vc_B)+D\partial^2_xc_B+\mu c_B(1-c_A-c_B)+
\sigma_B\xi'&
\end{eqnarray}
where now $<\xi(x,t)\xi'(x',t')>=0$,
$<\xi(x,t)\xi(x',t')>=\delta(t-t')\delta(x-x')$ and
$<\xi'(x,t)\xi'(x',t')>=\delta(t-t')\delta(x-x')$ . We also defined
$\sigma_A=\sqrt{\mu c_A(1+c_A+c_B)/N}$ and $\sigma_B=\sqrt{\mu
  c_B(1+c_A+c_B)/N}$.  We have chosen units of length such that
$L=1$. To return to dimensional units, simply let $1/N\rightarrow L/N$
in $\sigma_A$ and $\sigma_B$. To understand the connection with
stepping stone models, it is helpful to note that $L/N=\delta/N_l$, so
that the amplitude of the noise terms in Eqs.  (\ref{model1}) and
(\ref{model2}) can be written in a similar form as the ``generic
diffusion constant'' discussed in Ref. \cite{8}.
Eqs. (\ref{model1}) and (\ref{model2}) correspond to the macroscopic
equations presented in the Letter in absence of a selective advantage,
$s=0$. The derivation of the non-neutral is a straightforward
generalization in which the parameter $s$ controls the relative
difference of the duplication rate of species $A$ with respect to that
of species $B$. Mathematically, this correspondly to repeat the
expansion but substituting to the first equation in (\ref{mrates}) the
more general expression:
\begin{equation}\label{nonneutral}
W_A(+1,n_A,n_B)=\mu n_A\ \longrightarrow\ W_A(+1,n_A,n_B)=\mu(1+s) n_A
\end{equation}
leading to the the expression of Eq. (2) in the main text.

The continuous limit of such reaction-diffusion-advection processes is
only a concise summary of the stochastic dynamics; underlying this is
always the space-discretized version (see Ref. \cite{9}, page
314). One reason is that advection and diffusion terms can contribute
significantly to the variance of the noise when the cells are small
(\cite{9}, page 313); these terms can be neglected only if the
equations will be discretized on a sufficiently coarse
lattice. Moreover, the validity of the Kramers-Moyal (or Van Kampen)
expansion relies on $N_l\gg 1$. Even in absence of a carrying capacity
reduction due to compressible flows, the number of particles in a
sufficiently small cell may not be particularly large. The macroscopic
continuum equations will then be consistent with results from particle
simulations only when there is a sufficiently large number of
particles when we coarse-grain to scales small compared to the typical
scales of the density gradients. This is the case for the parameter
ranges considered in the main text.

\section{Continuous simulations and shell models}

In addition to particle simulations, Eq.s
(\ref{model1}-\ref{model2}) (i.e. Eqs. (2) in the main text) have been
simulated numerically by using finite difference methods in space and
the Euler-Cauchy method for time steps with a variety of velocity
fields $v(x,t)$. The effect of number fluctuations (i.e. genetic
drift) has been taken into account by using the method proposed by
Doering et. al. \cite{doering}: at each time step we compute the
position-dependent quantities $\sigma_{A,B} = \mu
c_{A,B}(1+c_A+c_B)/N_l$. If $\sigma_{A,B}>0$ (i.e. if there is a
non-zero number of particles of the appropriate type in the region of
size $\delta$), then we add the noise term; otherwise we set the noise
term equal to $0$. It has been shown by Doering et. al. that this
numerical scheme converges, in a suitable norm, to the continuum
limit.  In our numerical simulations, we found excellent agreement
between the results obtained by integrating eq.s
(\ref{model1}-\ref{model2}) with this technique, compared to those
obtained by particle methods, as will be discussed in details in a
forthcoming paper.

Next we illustrate how we built the "turbulent" velocity field
$u(x,t)$ used in our high Reynolds number simulations.  Although we
focus primarily on one spatial dimension, our analysis of the
statistical properties of $c(x,t)$ employs velocity fluctuations
typical of three dimensional turbulent fluid mechanics.  The
statistical properties of our advecting velocity field $u(x,t)$ will
be characterized by intermittency both in space and in time.  To
obtain a velocity field with generic space and time dynamics at high
Reynolds number, we build up $u(x,t)$ by appealing to a simplified
shell model of fluid turbulence \cite{biferale}. Wavenumber space is
divided into shells centered around $k_n = 2^{n-1} k_0$,
$n=1,2,...$. For each shell with characteristic wavenumber $k_n$, we
describe turbulence by using the complex Fourier-like variable
$u_n(t)$, satisfying the following equations of motion:
\begin{eqnarray}
  (\frac{d }{dt}&+&\nu k_n^2 ) u_n = i[k_{n+1} u_{n+1}^* u_{n+2}-\Delta
  k_n u_{n-1}^* u_{n+1} \nonumber  \\
&+&(1-\Delta) k_{n-1} u_{n-1} u_{n-2}]+f_n \ . 
\label{sabra}
\end{eqnarray}
The model contains one free parameter, $\Delta$, and it conserves two
quadratic invariants (when the force and the dissipation terms are
absent) for all values of $\Delta$. The first is the total energy
$\sum_n |u_n|^2$ and the second is $\sum_n (-1)^n k_n^{\alpha}
|u_n|^2$, where $\alpha= \log_{2} (1-\Delta)$.  We fix $\Delta =
-0.4$, as this choice reproduces intermittency features of the real
three dimensional Navier Stokes equation with surprising good accuracy
\cite{biferale}. Using $u_n$, we can build the real one dimensional
velocity field $u(x,t)$ as 
\begin{equation}
u(x,t) = F\sum_n [ u_n (t)e^{i k_n x} + u^*_n (t)e^{-i k_n x} ],
\end{equation}
where the constant $F$ controls the strength of velocity fluctuations
relative to other parameters in the model.  In all numerical
simulations we use a forcing function $f_n = [\epsilon(1+i)/u^{*}_1
]\delta_{n,1}$, i.e. energy is supplied only to the largest scale
corresponding to $n=1$, where $\epsilon$ is the injection rate of
turbulent energy.  With this choice, the power injected into the shell
model is simply given by $1/2\sum_n [u^{*}_n f_n + u_n f^{*}_n] =
\epsilon$ , i.e.  it is constant in time.  To solve
Eqs. (\ref{model1}-\ref{model2}) and (\ref{sabra}) we use a finite
difference scheme with periodic boundary conditions in space.

\section{Stochastic FKPP limit}

\begin{figure}[htb]
\includegraphics[width=8cm]{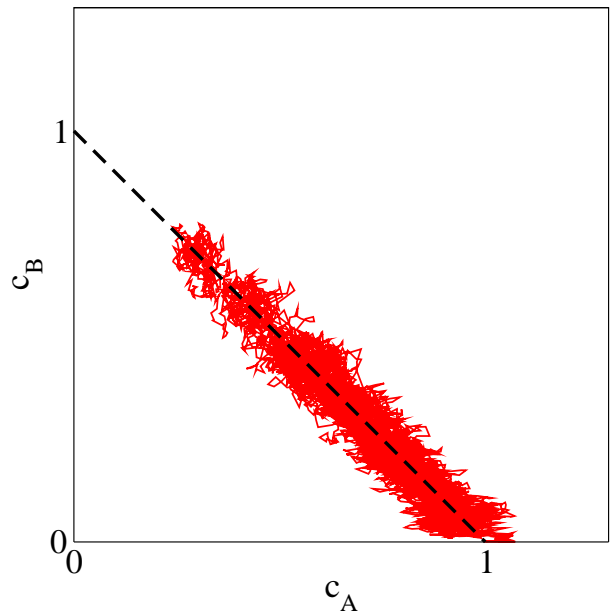}
\caption{Reaction plane. The dashed line is the steady state line with
  selective advantage $s=0$, $c_A+c_B=1$, where the model is
  equivalent to the stochastic FKPP equation \label{planefigure}. The
  trajectory is a zero-dimensional simulation of
  Eqs. (\ref{zerodmodel}) with $N=500$ and $\mu=1$. }
\end{figure}

In this section, we clarify the connection between the advection-free
model described in the Letter and in Eqs. (\ref{model1},\ref{model2})
and the stochastic FKPP (Fisher-Kolmogorov-Petrovsky-Piscounov)
equation.  To handle advecting flows, our model generalizates the FKPP
equation by relaxing the constraint on the total density,
$c_A(x,t)+c_B(x,t)=1$ for all values of $x$ and $t$, see
Fig. (\ref{planefigure}).

To understand this correspondence, it is helpful to change variables
from the densities $c_A$, $c_B$ to the relative densities
$f_{A,B}(x,t)=c_{A,B}(x,t)/c_T(x,t)$ where
$c_T(x,t)=c_A(x,t)+c_B(x,t)$. Eqs. (\ref{model1}-\ref{model2}) now become:
\begin{eqnarray}\label{stochfkpp}
  \partial_t f_A&=&-[v-D\partial_x\log(c_T)]\partial_x f_A+D\partial^2_x f_A
+\sigma_{f_A} \xi\nonumber\\
\partial_t f_B&=&-[v-D\partial_x\log(c_T)]\partial_x f_B+D\partial^2_x f_B
+\sigma_{f_B} \xi'
\end{eqnarray}
where now the noise amplitudes are given by
$\sigma^2_{f_A,f_B}=\mu(1+c_T)f_{A,B}(1-f_{A,B})/(Nc_T)$ and by
definition $f_A(x,t)+f_B(x,t)=1$.

Upon substituting $c_T=1$ and in absence of advection ($v=0$), each of
the fractions evolves according to the stochastic FKPP equations for
two neutral alleles. 

A further issue is to understand if there is a limit in which
the term $D\partial_x \log c_T$ can be neglected, so that one
retrieves Eq. (\ref{stochfkpp}) without having to impose any constraint.
An estimate follows from studying the fluctuations of
$c_T(x,t)=c_A(x,t)+c_B(x,t)$, which obeys a closed equation:
\begin{equation}
\partial_t c_T(x,t)=\mu c_T(1-c_T)+D\partial^2_xc_T
+\sqrt{\frac{\mu c_T(1+c_T)}{N}}\xi(x,t)
\end{equation}
where $\langle \xi(x,t)\xi(x't')=\delta(x-x')\delta(t-t')$. By
defining $\epsilon(x,t)=c_T(x,t)-1$ and assuming $\epsilon\ll1$, we
obtain a linear stochastic differential equation,
\begin{equation}
\partial_t \epsilon(x,t)=-\mu\epsilon+D\partial^2_x \epsilon
+\sqrt{\frac{2\mu}{N}}\xi(x,t)
\end{equation}

By means of a Fourier transform we can diagonalize the diffusion
operator and find that the Fourier modes $\tilde{\epsilon}(k,t)=\int\ dx
\exp(ikx) \epsilon(k,t)$ are independent Gaussian variables with zero mean
and variance $\sigma^2(k)=2\mu/[N(\mu+Dk^2)]$.
It follows that
\begin{equation}
\langle|\epsilon(x,t)|^2\rangle\approx\frac{2\mu}{N}\int_{-\infty}^{+\infty}
\frac{dk}{2\pi}\frac{1}{\mu+Dk^2}=\frac{1}{N}\sqrt{\frac{\mu}{D}}.
\end{equation}
The fluctuations in
$\epsilon(x,t)$ will be negligible when $N$ is large. In addition,
it is interesting to study
\begin{equation}
  \langle|D\partial_x \log c_T|^2\rangle \sim 
  \langle|D\partial_x \epsilon|^2\rangle = \int_{-k_{max}}^{+k_{max}}
  \frac{dk}{2\pi N}\  \frac{2\mu D^2k^2}{\mu+Dk^2}
\end{equation}
where we introduced a large $k$ cutoff (related to the
discretization size, $|k_{max}|\sim\delta^{-1}$) in
the above integral to avoid an ultraviolet divergence. With this
premise, the above integral is small if $D$ is small, suggesting that
the logarithmic terms in Eqs. (\ref{stochfkpp}) can be neglected in
this limit.

\end{document}